\documentclass[11pt, reqno, psamsfonts]{amsart}

\usepackage{latexsym, amsfonts, amsmath, amsthm}
\usepackage[psamsfonts]{amssymb}
\newtheorem{theorem}{Theorem}[section]
\newtheorem{lemma}[theorem]{Lemma}
\newtheorem{proposition}[theorem]{Proposition}
\newtheorem{corollary}[theorem]{Corollary}
\newtheorem{question}[theorem]{Question}

\theoremstyle{remark}
\newtheorem{remark}[theorem]{Remark}

\numberwithin{equation}{section}

\newcommand{\bj}{{\mathbf j}}

\newcommand{\Q}{{\mathbb Q}}

\newcommand{\N}{{\mathbb N}}

\newcommand{\CF}{\mathrm{CF}}
\newcommand{\PP}{\mathrm{P}}
\newcommand{\NP}{\mathrm{NP}}
\newcommand{\VP}{\mathrm{VP}}
\newcommand{\VNP}{\mathrm{VNP}}

\newcommand{\al}{\alpha}
\newcommand{\be}{\beta}
\newcommand{\ga}{\gamma}
\newcommand{\la}{\lambda}
\newcommand{\sig}{\sigma}

\newcommand{\cO}{{\mathcal O}}
\newcommand{\cR}{{\mathcal R}}
\newcommand{\cS}{{\mathcal S}}

\newcommand{\ran}{{\rangle}}
\newcommand{\lan}{{\langle}}
\newcommand{\eps}{\epsilon}

\newcommand{\Per}{\mathrm{PER}}
\newcommand{\Imm}{\mathrm{IM}}
\newcommand{\imm}{\mathrm{im}}
\newcommand{\per}{\mathrm{per}}
\newcommand{\Det}{\mathrm{DET}}
\newcommand{\Ind}{\mathrm{Ind}}

\newcommand{\lambn}{\la^{(n)}}
\newcommand{\Imlan}{\Imm_{\lambn}}

\begin{document}
\date{January 23, 2003}

\title
{Complexity and Completeness of  immanants}

\author{Jean-Luc Brylinski}
\address{JLB: P.O. Box 1089, Truro,  MA 02666}
\email{jlb496@yahoo.com}

\author{Ranee Brylinski}
\address{RKB: Department of Mathematics,
        Penn State University, University Park 16802\newline
  \phantom{aaaa}      Current address: P.O. Box 1089, Truro,
  MA 02666}
\email{rkb@math.psu.edu, rkb248@yahoo.com}
\urladdr{www.math.psu.edu/rkb}


\begin{abstract}
Immanants are polynomial functions of $n$ by $n$  matrices
attached to irreducible  characters of the symmetric group
$\cS_n$, or equivalently to Young diagrams of size  $n$. Immanants
include determinants and permanents as extreme cases. Valiant
proved that computation of  permanents is a complete problem in his
algebraic model of $\NP$ theory, i.e., it is $\VNP$-complete.
We prove that computation of  immanants  is $\VNP$-complete if
the immanants are attached to a family of
diagrams whose separation is
$\Omega(n^\delta)$ for some $\delta>0$. We define the separation of a
diagram to be  the largest number of overhanging boxes contained in a
single row. Our theorem proves a conjecture of B\"urgisser for a
large variety of families, and in particular we recover with new
proofs his $\VNP$-completeness results for hooks and rectangles.
\end{abstract}

\maketitle

\section{Introduction} \label{sec:intro}

In algebraic complexity theory, one considers families $(f_n)$ of multivariate
 polynomials, where both the number of variables and the degree
 are polynomially bounded functions of $n$
 (i.e., are of the form  $O(p(n))$ for some
 polynomials $p(n)$).  The complexity of
    $(f_n)$ is the minimum possible size (or cost) of a computation of $(f_n)$.
     This is a function of $n$, and the growth of this function is
     what matters.
Of course, we must first choose a suitable computational model for
the $f_n$. In this context, a computational model is some type of
arithmetic circuit. We assume the  circuit inputs the variables
and some scalars, and then computes $f_n$ by performing the
arithmetic operations $+,-,*$ on the inputs and previously
computed quantities.    The \textit{size} of this circuit is then
the number of operations used to compute $f_n$. The size of the
smallest possible circuit is called the
\textit{\textup{(}total\textup{)} algebraic complexity} $L(f_n)$.

In this theory,  the determinant and the permanent play a special
role. The determinant and permanent families $(\Det_n)$ and
$(\Per_n)$ are defined by
\begin{equation}\label{eq:Det+Pert}
\Det_n=\sum_{\pi\in\cS_n} \eps(\pi) \prod_{i=1}^n X_{i,\pi(i)}
\qquad\textrm{ and }\qquad \Per_n=\sum_{\pi\in\cS_n} \prod_{i=1}^n
X_{i,\pi(i)}\end{equation}
 where $X=(X_{ij})$ is an $n$ by $n$
matrix, $\cS_n$ is the symmetric group in $n$ letters, and
$\eps(\pi)$ is the sign of a permutation $\pi$. Although
$(\Det_n)$ and $(\Per_n)$ look very similar, their complexities
are (apparently) very different. The determinant family is
\textit{easy} to compute, in the sense that its algebraic
complexity is polynomially bounded. In fact, $O(n^4)$ operations
are enough to compute $\Det_n$; see e.g. \cite{bcss}. In Valiant's
algebraic model of $\PP-\NP$ theory, $(\Det_n)$ is the analog of a
$\PP$ decision problem.

However, the permanent family is apparently \textit{hard} to
compute, in that no known polynomial size circuit computes
$(\Per_n)$. The smallest known circuits for $(\Per_n)$ require
$O(n\cdot 2^n)$ arithmetic operations \cite{rys, nw}. In Valiant's
theory, $(\Per_n)$ is the algebraic analog of an $\NP$-complete
decision problem (assuming the characteristic of the ground field
is different from $2$). Precisely, Valiant discovered \cite{V1,V3,
vzG} that the permanent family is $\VNP$-complete. This was a hard
and surprising result. Valiant's hypothesis is true if and only if
the algebraic complexity of $(\Per_n)$ truly grows faster than any
polynomial function of $n$.

The representation theory of the symmetric group provides a
natural way to construct families which are intermediate between
the determinants and the permanents. The new families consist of
immanants.  For each irreducible character $\chi_\la$ of the
symmetric group $\cS_n$, Littlewood \cite{li} defined the immanant
of $X$ to be the polynomial
\begin{equation}\label{eq:imm}
\Imm_\la=\sum_{\pi\in\cS_n}\chi_\la(\pi) \prod_{i=1}^n
X_{i,\pi(i)}
\end{equation}
Here $\la=(\la_1,\dots,\la_\ell)$ is a partition of size $n$. So
$\la_1\ge\cdots\ge \la_\ell>0$ are  integers, $\ell=\ell(\la)$ is
the \textit{length} of $\la$ and $|\la|=\la_1+\cdots+\la_\ell$ is
the \textit{size} of $\la$. We set $\la_{\ell+1}=0$. If
$\la=(1^n)$, then $\chi_\la$ is the sign character, and so
$\Imm_\la=\Det_n$. If $\la=(n)$, then $\chi_\la$ is the trivial
character and so $\Imm_\la=\Per_n$. We can identify $\la$ with its
Young diagram. This is a left-justified array of $\ell$ rows of
boxes, with exactly $\la_i$ boxes in the $i$-th row. So $\Det_n$
and $\Per_n$ correspond to the two extreme partitions  where $\la$
is simply a column or a row. The \textit{width} of $\la$ is then
the size $\la_1$ of its largest row.

Lower bound results for the complexity of  immanants have been
found by Hartmann \cite{ha}, and most recently, by B\"urgisser
\cite{bur1, bur_book}. One main question here is to figure out
when an immanant family is $\VNP$-complete, i.e., has the same
hardness as the permanent family. B\"urgisser showed (see
Proposition \ref{prop:Im_is_vnp})  that for this it is natural to
consider a family $(\lambn)$ of partitions where the size
$|\lambn|$ is polynomially bounded in $n$. B\"urgisser conjectured
that if also the width of $\lambn$ is $\Omega(n^\delta)$ for some
$\delta>0$, then the family $(\Imlan)$ is $\VNP$-complete.
B\"urgisser's main result in \cite{bur1} is the proof of his
conjecture for two types of families of partitions, one where the
shapes were all hooks $(n-i,1^{i})$, and the other where the
shapes were all rectangles $(m^{s})$.

In this paper, we prove  B\"urgisser's conjecture for a large
variety of families, where we replace the width of each partition
$\la$ by the parameter
\[  k=\max\nolimits_{i=1}^{\ell}\{\la_i-\la_{i+1}\} \]
We call $k$ the \textit{separation} of $\la$. For instance, the
separation of the hook $(h,1^i)$, $h\ge 2$,  is $h-1$ and that of
the rectangle $(m^s)$ is $m$. Our main result is

\begin{theorem}\label{thm:VNP_compl}
Let $(\lambn)$ be a family of partitions such that
\begin{itemize}
\item[(i)] the size $|\lambn|$ is  polynomially bounded
in $n$;
\item[(ii)] the separation $k(n)$ of $\lambn$ satisfies
$k(n)=\Omega(n^\delta)$ for some $\delta>0$.
\end{itemize}
Then the corresponding immanant family $(\Imlan)$  is
$\VNP$-complete.
\end{theorem}

This result recovers, with new proofs, the two cases of hooks and
rectangles treated by B\"urgisser. (However in the case of hook
shapes, we do not recover his result on $\#\PP$-completeness).

We prove Theorem \ref{thm:VNP_compl} in a rather simple way, by
constructing an explicit projection, in the sense of Valiant, from
$\Imm_\la$ to $\Per_k$, where $k$ is the separation of $\la$ (in
fact, for any  $k\in\{\la_i-\la_{i+1}\}_{i=1}^\ell$).
Constructing such a projection
means the following. For any $k$ by $k$ matrix $A$, we construct
an $n$ by $n$ matrix $A^\sharp$ such that \textup{(i)} The value
of $\Imm_\la$ at $A^\sharp$ is equal to the value of $\Per_k$ at
$A$ and \textup{(ii)} each entry of $A^\sharp$ is either a scalar
or an entry of $A$. What makes our proof simple is the nature of
$A^\sharp$. Our matrix $A^\sharp$ is block diagonal. The first
block is $A$ and the subsequent blocks are scalar matrices, drawn
from a list $\{ H_1,E_1,H_2,E_2,\dots\}$ where $H_q$ and $E_q$ are
some explicit $q$ by $q$ matrices; see Proposition \ref{prop:proj}
and Lemma \ref{lem:HandE}.

Here is the organization of the paper. In Section 2, we recall
some key notions of the Valiant's  theory for families of
polynomials.   We also state B\"urgisser's result on immanant
families lying in $\VNP$.

In Section 3, we explain  our   projection results. The main
result, Proposition \ref{prop:proj}, is somewhat abstract, but it
easily leads to Corollary \ref{cor:proj}, in which we project an
immanant to a permanent. Sections
\ref{sec:blockdiag}-\ref{sec:proof_prop_proj}  are devoted to
proving Proposition \ref{prop:proj}. Finally, in Section
\ref{sec:thm1.1} we complete the proof of Theorem
\ref{thm:VNP_compl}.

Returning to B\"urgisser's conjecture, we note that our results in
this paper give no useful information in the case where the
separation of $\lambn$ grows too slowly for Theorem
\ref{thm:VNP_compl} to apply. We expect the key question here is
\begin{question} Suppose $\lambn$ is the staircase partition
$(n,n-1,\dots,2,1)$. Is the corresponding immanant family
$\VNP$-complete?
\end{question}
An affirmative answer to this question, together with a reasonable
explanation, should lead to a proof of B\"urgisser's conjecture. A
negative answer would of course disprove B\"urgisser's conjecture.

\section{Immanant Families and Valiant's algebraic model of
$\PP-\NP$ theory} \label{sec:valiant}

In this section, we recall some definitions and results concerning
Valiant's complexity classes. See \cite{V1, V3, vzG, str, bcs,
bur_book} for more information.

We fix a field $k$ of characteristic $0$; in particular $k$ can be
the field $\Q$ of rational numbers.  For a multivariate polynomial
$f\in k[X_1,\dots,X_v]$,   the \textit{total algebraic complexity}
$L(f)$ of $f$ is the minimum number of arithmetic operations
$+,-,*$ needed to compute $f$ in an arithmetic circuit (or
straight line program), using only inputs in $k\cup
\{X_1,\dots,X_v\}$. For instance, if $f(X)=X^{2^n}$, then
$L(f)=n$.

 A \textit{$p$-family} $(f_n)$ is a sequence
of multivariate polynomials $f_n\in k[X_1,\dots,X_{v(n)}]$ such
that both the number $v(n)$ and the degree $deg(f_n)$ are
polynomially bounded. The determinant and permanent families
$(\Det_n)$ and $(\Per_n)$ discussed in  Section \ref{sec:intro}
are $p$-families. Clearly, $(\Imlan)$ is a $p$-family if and only
if the size $|\lambn|$ is polynomially bounded.

A $p$-family $(f_n)$ is called \textit{$p$-computable} if the
complexity $L(f_n)$ is polynomially bounded. The set of all
$p$-computable families $(f_n)$ is Valiant's complexity class
$\VP$. For instance, $(\Det_n)$  lies in $\VP$ since
$L(\Det_n)=O(n^4)$.

A family $(f_n)$, $f_n\in k[X_1,\dots,X_{v(n)}]$, is called
\textit{$p$-definable} if there exists a $p$-computable family
$(g_n)$, $g_n\in k[X_1,\dots,X_{u(n)}]$, with $u(n)\ge v(n)$ such
that
\begin{equation}
  f_n(X_1,\dots,X_{v(n)})=\sum_{e\in \{ 0,1\}^{u(n)-v(n)}}~
g_n(X_1,\dots,X_{v(n)},e_{v(n)+1},\dots,e_{u(n)})
\end{equation}
The set of $p$-definable families is Valiant's complexity class
$\VNP$.  It is not hard to show that $(\Per_n)$ is $p$-definable.
B\"urgisser proved that every $p$-family of immanants is
$p$-definable.

\begin{proposition}[{\cite[Prop. 4.1]{bur1},
\cite[Prop. 7.9, p. 126]{bur_book}}] \label{prop:Im_is_vnp} If
$(\lambn)$ is a sequence of partitions such that the size
$|\lambn|$ is polynomially bounded, then the corresponding
immanant family $(\Imlan)$ belongs to $\VNP$.
\end{proposition}

Valiant's notion of $\VNP$-completeness is based on the following
simple notion of reduction. A polynomial $f\in k[X_1,\dots,X_v]$
is  called a {\it projection} of a polynomial $g\in
k[X_1,\dots,X_u]$ if for some $a_1,\dots,a_u$ lying in $k\cup \{
X_1,\dots,X_v\}$ we have the identity
\begin{equation}
f(X_1,\dots,X_v)=g(a_1,\dots,a_u)
\end{equation}
 In this case, we write $f\leq g$. For
instance, if $f_k(X_1,\dots,X_k)=X_1\dots X_k$ and $n\ge k$, then
$f_k\leq \Det_n$ and $f_k\leq \Per_n$. This follows by
specializing  $X=(X_{ij})$ to an appropriate diagonal matrix. Note
that $f\leq g$ implies $L(f)\leq L(g)$.

A $p$-family $(f_n)$ is  a \textit{projection} of the $p$-family
$(g_n)$ if there exists a function $t:\N\to\N$ such that $t$ is
polynomially bounded  and $f_n\leq g_{t(n)}$ for all $n$. A
$p$-definable family $(f_n)$ is called \textit{$\VNP$-complete} if
every family $(g_n)\in \VNP$ is a projection of $(f_n)$. We have

\begin{theorem}[Valiant \cite{V1, vzG}] \label{thm:per}
The permanent family $(\Per_n)$ is $\VNP$-complete.
\end{theorem}
This result is surprising since $(\Per_n)$ is the enumerator for
the  problem of deciding if  a bipartite graph  has a perfect
matching, and this decision problem belongs to $\PP$. An immediate
consequence is

\begin{corollary}\label{cor:VNP_criterion}
Suppose $(f_n)$, $f_n\in k[X_1,\dots,X_{v(n)}]$, is a
$p$-definable family.   Then $(f_n)$ is $\VNP$-complete if and
only if $(f_n)$ projects to $(\Per_n)$.
\end{corollary}

Valiant's hypothesis asserts $\VP\neq\VNP$. This is a (nonuniform)
algebraic analog of the famous Cook hypothesis $\PP\neq\NP$.

\section{Projection formulas for immanants}\label{sec:proj}

In this section, we describe our projection results. For this
purpose, we use  (as in \cite{bcs, bur_book, bur1}) the following
notation. If $A$ is a $d$ by $d$ matrix and $|\la|=d$, then
$\imm_\la(A)$ and $\per(A)$ denote the values of $\Imm_\la$ and
$\Per_d$ at $A$.

Our key result is Proposition \ref{prop:proj} below. To state
this, we introduce the diagonal $q$ by $q$ matrix $D_q$
\begin{equation}
D_q=\left(\begin{array}{cccc}
  1& 0&\cdots& 0\\
    0& \frac{1}{2}&\cdots&0\\
\vdots &\vdots &&\vdots\\
0& 0&\cdots& \frac{1}{q}
  \end{array}\right)
  \end{equation}
with entries $1,\frac{1}{2},\cdots,\frac{1}{q}$.

We also need two notions concerning partitions. A \textit{
horizontal strip} inside $\la$ is a set $S$ of boxes of $\la$ such
that if $s$ lies in $S$, then (a) all boxes to the right of $s$ in
the same row also belong to $S$, and (b) $s$ is the lowest box in
its column. Similarly, one also defines a  \textit{ vertical
strip} inside $\la$. In both cases, the size of the strip $S$ is
the total number of boxes in $S$. Notice that if we remove from
$\la$ either a horizontal strip or a vertical
 strip, then what remains is again a
partition.

  \begin{proposition} \label{prop:proj}
  Let $\la$ be a partition of $n$, and choose $q\leq
  n$. We have:

\begin{itemize}
\item[(i)]
  $\Imm_\la$ projects to $\sum_\mu \Imm_\mu$, where the sum is over
  all partitions $\mu$ obtained by removing from  $\la$  a
  horizontal strip of size $q$.
\item[(ii)] $\Imm_\la$ projects to $\sum_\nu \Imm_\nu$, where the sum is over
  all partitions obtained by removing from  $\la$  a
  vertical strip of size $q$.
\end{itemize}

  We can realize the projections explicitly in the following way,
  where $A$ is a square matrix of size $n-q$.
  In \textup{(i)} we have
\begin{equation}\label{eq:proj_row}
  \sum_\mu \imm_\mu (A)=\imm_\la\left(\begin{array}{ll} A& 0\\ 0&H_q
  \end{array}\right)
\end{equation}
where $H_q$ is the $q$ by $q$ matrix

\begin{equation}\label{eq:Hq}
  H_q=D_q\left(\begin{array}{lll}1&\cdots&1\\ \vdots&&\vdots\\
1&\cdots&1\end{array}\right)=\left(\begin{array}{lll}1&\cdots&1\\
\frac{1}{2}&\cdots&\frac{1}{2}\\ \vdots&&\vdots\\
\frac{1}{q}&\cdots&\frac{1}{q}\end{array}\right)
\end{equation}

In \textup{(ii)} we have

\begin{equation}\label{eq:proj_column}
  \sum_\nu \imm_\nu (A)=\imm_\la\left(\begin{array}{ll} A& 0\\ 0&E_q
  \end{array}\right)
\end{equation}
where $E_q$ is the $q$ by $q$ matrix

\begin{equation}\label{eq:Eq}
  E_q=D_q\left(\begin{array}{ccccc} 1&1&0&\cdots&0\\
-1&1&2&\cdots&0\\
\vdots&\vdots&\vdots&&\vdots\\
(-1)^{q-2}&(-1)^{q-3}&\cdot&\cdots&q-1\\
(-1)^{q-1}&(-1)^{q-2}&\cdot&\cdots&1
\end{array}\right)
\end{equation}

\end{proposition}

We prove Proposition \ref{prop:proj} in Sections
\ref{sec:blockdiag}-\ref{sec:proof_prop_proj}. In this paper, we
will only use the following corollary:

\begin{corollary} \label{cor:proj}
Let $\la$ be a partition of $n$. Then
\begin{itemize}
\item[(i)] $\Imm_\la$ projects to $\Imm_\mu$, where $\mu$ is obtained
from $\la$  by removing its first row.

\item[(ii)] $\Imm_\la$ projects to $\Imm_\nu$, where $\nu$ is obtained
from $\la$  by removing its first column.
\end{itemize}

The explicit formulas are as in Proposition
\textup{\ref{prop:proj}}, where in \textup{(i)} we choose $q$ to
be the size of the first row of $\la$, and in \textup{(ii)} we
choose $q$ to be the size of the first column of $\la$.
\end{corollary}

\begin{proof}
We will prove (i); the proof of (ii) is entirely similar. Let $q$
be the size of the first row of $\la$; so $q=\la_1$. Then there is
a unique horizontal strip of size $q$ inside $\la$: this contains
the lowest box in each column. Removing this strip from $\la$ has
the effect of  shortening the $i$th row  from $\la_i$ to
$\la_{i+1}$. The remaining partition is then
$\mu=(\la_2,\dots,\la_\ell)$. Thus $\Imm_{\la}$ projects to
$\Imm_{\mu}$ by Proposition \ref{prop:proj}(i).
\end{proof}

By making successive applications of Corollary \ref{cor:proj}, we
can project $\Imm_{\la}$ to $\Per_k$ for any
$k\in\{\la_i-\la_{i+1}\}_{i=1}^\ell$. This is because, by
successively removing rows and columns from $\la$, we can obtain
the row partition $(k)$. For example, if $\la$ is the hook
partition $(n-i,1^i)$, then removing the first column of $\la$
leaves the row $(n-i-1)$. So here $\Imm_\la$ projects to
$\Per_{n-i-1}$. If $\la$ is the rectangle $(m^s)$, then removing
the first $s-1$ rows of $\la$ leaves the row $(m)$. So then
$\Imm_\la$ projects to $\Per_m$.

More generally, given an arbitrary partition $\la$, we can remove
the first $i-1$ rows of $\la$. This leaves the partition
$\la^\sharp=(\la_i,\dots,\la_{\ell})$. Then we can remove the
first $\la_{i+1}$ columns of $\la^\sharp$; these have lengths
$\pi_1,\dots,\pi_{\la_{i+1}}$, where $\pi$ is the conjugate
partition to $\la^\sharp$. This leaves exactly the row
$(\la_i-\la_{i+1})$.  Then for any square matrix $A$ of size
$\la_i-\la_{i+1}$ we have
\begin{equation}\label{eq:perkA}
  \per_{\la_i-\la_{i+1}}(A)=\imm_\la(G)
\end{equation}
where
\begin{equation}\label{eq:G}
  G=diag(A,H_{\la_1},\dots,H_{\la_{i-1}},E_{\pi_1},
  \dots,E_{\pi_{\la_{i+1}}})
\end{equation}
is the block diagonal matrix made up of the indicated blocks.
Finally, we can chose $i$ so that we maximize $\la_i-\la_{i+1}$.
Thus we get

\begin{corollary} $\Imm_\la$ projects to $\Per_k$ where
$k=\max\nolimits_{i=1}^{\ell}\{\la_i-\la_{i+1}\}$ is the
separation of $\la$. We can realize the projection explicitly as
in \textup{(\ref{eq:perkA})} and \textup{(\ref{eq:G})}.
\end{corollary}

\section{Computing immanants of
block diagonal matrices }\label{sec:blockdiag}

To prove Proposition \ref{prop:proj}, we start with the following
simple observation about the immanants of the
block diagonal matrix
$\bigl(\begin{smallmatrix}A&0\\ 0&B\end{smallmatrix}\bigr)$.
Let $V_\la$ be the irreducible representation of $\cS_n$ with
character $\chi_\la$.

\begin{lemma} \label{lem:perm_block} Let $\la$ be a partition of $n$
 and write $n=p+q$.
Suppose $A$ is a $p$ by $p$ matrix and $B$ is a $q$ by $q$ matrix.
Then

\begin{equation}
\imm_\la \left( \begin{array} {ll}A&0 \\ 0&B\end{array}\right)=
\sum_{|\al|=p,|\be|=q} c^\la_{\al,\be}\imm_\al (A)\imm_\be (B)
\end{equation}

\noindent where $c^\la_{\al,\be}$ is the multiplicity of $V_\la$
in the induced representation
$\Ind_{\cS_p\times\cS_q}^{\cS_{p+q}}~V_\al\otimes V_\be$.
\end{lemma}

Here $\cS_p\times\cS_q$ is the subgroup of $\cS_{p+q}$ formed in
the usual way. I.e., we can represent $\cS_{p+q}$ as the
permutations of $\{ 1,2,\dots,n\}$ and then $\cS_p\times\cS_q$ is
the set of elements $\sig\tau$ where $\sig$ is a permutation of
$\{ 1,2,\dots,p\}$ and $\tau$ is a permutation of $\{
p+1,p+2,\dots,n\}$.

\begin{proof}
Let $M=\bigl(\begin{smallmatrix}A&0\\ 0&B\end{smallmatrix}\bigr)$.
We have $\imm_\la (M)=\sum_{\pi\in\cS_n}\chi_\la(\pi)f_\pi(M)$
where $f_\pi(M)=\prod_{i=1}^n M_{i,\pi(i)}$. Clearly, $f_\pi(M)$
vanishes unless $\pi$ belongs to the subgroup $\cS_p\times\cS_q$.
So we assume $\pi=\sig\tau$ where $\sig\in\cS_p$ and
$\tau\in\cS_q$.  Then $f_\pi(M)=f_\sig(A)f_\tau (B)$, where we
define $f_\sig$ and $f_\tau$ in the same way as $f_\pi$. Our aim
now is to compute the character values $\chi_\la(\sig\tau)$. But
$\chi_\la(\sig\tau)$ is the trace of $\sig\tau$ on $V_\la$, and so
we need to decompose $V_\la$ as a representation of
$\cS_p\times\cS_q$. By Frobenius reciprocity, this decomposition
is
\begin{equation}
V_\la \downarrow \cS_p\times\cS_q=\bigoplus_{|\al|= p,|\be|= q}
c^{\la}_{\al,\be} V_\al\otimes V_\be
\end{equation}
where the  coefficients $c^{\la}_{\al,\be}$ were defined in the
statement of the lemma. Thus $\chi_\la(\sig\tau)=\sum_{\al,\be}
c^{\la}_{\al,\be} \chi_\al(\sig)\chi_\be(\tau)$. Then
\begin{equation}
\imm_\la(M)=\sum_{\sig,\tau,\al,\be}c^{\la}_{\al,\be}\chi_\al(\sig)\chi_\be(\tau)
f_\sig(A)f_\tau(B)=\sum_{\al,\be}c^{\la}_{\al,\be}\imm_\al(A)\imm_\be(B)
\end{equation}

\end{proof}

\section{The matrices $H_q$ and $E_q$}
\label{sec:HqandEq}

To use Lemma \ref{lem:perm_block} to prove Proposition
\ref{prop:proj} we need the following properties of $H_q$ and
$E_q$.

\begin{lemma} \label{lem:HandE}
\begin{itemize}
\item[(i)] $\per(H_q)=1$ while $\imm_\be(H_q)=0$ for all
$\be$  different from $(q)$.
\item[(ii)] $\det(E_q)=1$ while $\imm_\be(E_q)=0$ for all
$\be$  different from $(1^q)$.
\end{itemize}
\end{lemma}

\begin{proof}
(i) Let $R_q$ be the $q$ by $q$ matrix with all entries equal to
$1$, so that $H_q=D_qR_q$. Then
$\imm_\be(H_q)=\frac{1}{q!}\imm_\be(R_q)$ for each partition $\be$
of $q$, and we get

\begin{equation}
\imm_\be(H_q)=\frac{1}{q!}\sum_{\pi\in\cS_q} \chi_\be(\pi)=
\frac{1}{q!}\lan \chi_\be,\chi_{(q)} \ran \end{equation}
 where
$\lan ~,~\ran$ is the usual inner product of characters. We know
the irreducible characters of $\cS_q$ are orthogonal and
$\lan\chi_{(q)},\chi_{(q)} \ran=q!$. So
$\per(H_q)=\imm_{(q)}(H_q)=1$ and $\imm_\be(H_q)=0$ if $\be\neq
(q)$.

(ii) We have $E_q=D_qT_q$ where $T_q$ is the last matrix in
(\ref{eq:Eq}). Then $\imm_\be(E_q)=\frac{1}{q!}\imm_\be(T_q)$ for
each partition $\be$ of $q$. Littlewood \cite[pp. 83-86]{li}
introduced the matrix
\begin{equation}\label{eq:Z}
  Z=\left(\begin{array}{ccccc} \zeta_1&1&0&\cdots&0\\
\zeta_2&\zeta_1&2&\cdots&0\\
\vdots&\vdots&\vdots&&\vdots\\
\zeta_{q-1}&\zeta_{q-2}&\cdot&\cdots&q-1\\
\zeta_{q}&\zeta_{q-1}&\cdot&\cdots&\zeta_1
\end{array}\right)
\end{equation}
where $\zeta_1,\dots,\zeta_q$ are indeterminates, and he proved
the formula
\begin{equation}\label{eq:imZ}
\imm_\be(Z)= \sum_{|\ga|=q}d_\ga\chi^\ga_\be \zeta^\ga
\end{equation}
where $\zeta^\ga=\zeta_{\ga_1}\cdots \zeta_{\ga_r}$ if
$\ga=(\ga_1,\dots,\ga_r)$ with $\ell(\ga)=r$, and $d_\ga$ is the
number of permutations of cycle type $\ga$. Here $\chi^\ga_\be$ is
the value $\chi_\be(\pi)$ for any $\pi\in\cS_q$ of cycle type
$\ga$. Now we obtain the immanants of $T_q$ by specializing the
$\zeta_i$ so that $\zeta_i=(-1)^{i+1}$. Notice that $\zeta^\ga$
specializes to  $\eps(\ga)$ where $\eps$ is the sign character and
$\eps(\ga)$ is the sign of any permutation of cycle type $\ga$. So
we find
\begin{equation}
  \imm_\be(E_q)=\frac{1}{q!}\imm_\be(T_q)=
  \frac{1}{q!}\sum_{|\ga|=q}d_\ga\chi^\ga_\be\eps(\ga)=
  \frac{1}{q!}\sum_{\pi\in\cS_q}\chi_\be(\pi)\eps(\pi)=
  \frac{1}{q!}\lan \chi_\be,\eps\ran
\end{equation}
By orthogonality of characters again, we know
$\lan\chi_\be,\eps\ran$ is zero unless $\chi_\be=\eps$. This
happens when $\be=(1^q)$ and then $\lan\eps,\eps\ran=q!$. So
$\det(E_q)=\imm_{(1^q)}(E_q)=1$ and $\imm_\be(E_q)=0$ if $\be\neq
(1^q)$.

Finally, for completeness, we recall Littlewood's proof of
(\ref{eq:imZ}). Let $\cO_\ga$ be the set of permutations of cycle
 type $\ga$, so that  $d_\ga=|\cO_\ga|$. Then
it suffices to compute
 the cycle format polynomials
 $\CF_\ga(Z)=\sum_{\pi\in\cO_\ga}f_\pi(Z)$
 (where   $f_\pi$ was defined in the proof
 of Lemma \ref{lem:perm_block}), since
$\imm_\be(Z)=\sum_{|\ga|=q}\chi^\ga_\be\CF_\ga(Z)$. We will show
that
\begin{equation}\label{eq:CF_ga}
  \CF_\ga=d_\ga\zeta^\ga
\end{equation}
Certainly (\ref{eq:CF_ga}) implies (\ref{eq:imZ}). (In fact,
 they  are equivalent).

 To compute $\CF_\ga(Z)$, we first observe that
  $f_\pi(Z)$ is non-zero only if
 each cycle of $\pi$  is of the form
 $\theta_{i+s,s}=(i+1,i+2,\dots,i+s)$ for some $i$ and $s$.
 Then  $\theta_{i+s,s}$  contributes
 the factor $(i+1)\cdots (i+s-1)\zeta_s$ to $f_\pi(Z)$.

Thus the $\pi$ in $\cO_\ga$ with
 $f_\pi(Z)\neq 0$ are all obtained in the following way:
 we take a permutation $\bj=(j_1,\dots,j_r)$ of
 $(\ga_1,\dots,\ga_r)$ and then $\pi$ is the product
 of the cycles
 $\theta_{j_1,j_1}\theta_{j_1+j_2,j_2}\cdots\theta_{j_1+\dots+j_r,j_r}$.
 Then
\begin{equation}\label{eq:fpiZ}
  f_\pi(Z)=1\cdot 2\cdots (j_1-1)\zeta_{j_1}(j_1+1)\cdots
  (j_1+j_2-1)\zeta_{j_2}(j_1+j_2+1)\cdots=\frac{q!}{\phi(\bj)}
  \zeta^\ga
\end{equation}
where $\phi(\bj)=j_1(j_1+j_2)\cdots (j_1+\cdots+j_r)$. So
$\CF_\ga(Z)=(\sum_\bj q!/\phi(\bj))\zeta^\ga$ where we sum over
all distinct choices for $\bj$. Fortunately, we can recognize
$\sum_\bj q!/\phi(\bj)$ as the size $d_\ga$ of $\cO_\ga$.
Littlewood proved this numerically by induction on $r$ (using the
formula $d_\ga=q!/1^{m_1}m_1!2^{m_2}m_2!\cdots$ where
$\ga=(1^{m_1}2^{m_2}\cdots)$) but instead we will prove it by a
counting argument where we partition $\cO_\ga$ into subsets.

We attach to each $\pi\in\cO_\ga$ a sequence
$\bj=(j_1,\cdots,j_r)$ in the following way. Given a cycle
$\sig=(t_1,\dots,t_s)$ in $\cS_q$, we put
$\max(\sig)=\max\{t_1,\dots,t_s\}$. Clearly there is  a unique way
to write $\pi=\sig_1\cdots\sig_r$ as a product of $r$ disjoint
cycles so that $\max(\sig_1)<\cdots <\max(\sig_r)$. Now let $j_i$
be the length of $\sig_i$. Then $\bj$ is a permutation of
$(\ga_1,\cdots,\ga_r)$. Let $\cO_\ga(\bj)$ be the set of $\pi$
attached to $\bj$ in this way. The cardinality of $\cO_\ga(\bj)$
is exactly $q!/\phi(\bj)$. To see this, we can associate to
$\pi=\sig_1\cdots\sig_r$ the word $w(\pi)=w_1\cdots w_q$ where
$(w_{j_1+\cdots+j_{i-1}+1},\dots,w_{j_1+\cdots+j_i})=\sig_i$ and
$w_{j_1+\cdots+j_i}=\max(\sig_i)$.  Notice that $w(\pi)$ lies in
$\cS_q$, i.e., is just a permutation of the word $12\cdots q$.
This sets up a bijection between $\cO_\ga(\bj)$ and the set of
words $w$ such that $w\in\cS_q$  and $w_k=\max\{ w_1,\dots,w_k\}$
if $k\in \{ j_1,j_1+j_2,\dots,j_1+\cdots+j_r\}$. Clearly
$1/(j_1+\cdots+j_r)$ of all words $u\in\cS_q$ satisfy $\max\{
u_1,\dots,u_{j_1+\cdots+j_r}\}=u_{j_1+\cdots+j_r}$, and
$1/(j_1+\cdots+j_{r-1})$ of these words satisfy $\max\{
u_1,\dots,u_{j_1+\cdots+j_{r-1}}\}=u_{j_1+\cdots+j_{r-1}}$, and so
on. Thus $q!/\phi(\bj)=|\cO_\ga(\bj)|$.
 We have the disjoint union $\cO_\ga=\cup_\bj \cO_\ga(\bj)$ and
so we conclude $d_\ga=\sum_\bj q!/\phi(\bj)$. This proves
(\ref{eq:CF_ga}) and hence also (\ref{eq:imZ}).
\end{proof}

\begin{remark}\label{rem:Littlewood}
Littlewood actually discussed the identity (\ref{eq:imZ}) in the
setting where $\zeta_i$ is the $i$-th power sum symmetric function
$p_i=p_i(x_1,\cdots,x_n)=x_1^i+\cdots+x_n^i$ in indeterminates
$x_1,\dots,x_n$. (The power sums $p_1,\dots,p_q$ are algebraically
independent as long as $q\leq n$, and so there was no harm in
thinking of them as indeterminates). Littlewood's result
(\ref{eq:imZ}) was then that $\imm_\be(Z)$ is $q!$ times the Schur
function $s_\be(x_1,\dots,x_n)$.
\end{remark}

\begin{remark}
In (i) and (ii) of Lemma \ref{lem:HandE}, and of Proposition
\ref{prop:proj}, there are other choices for $H_q$ and $E_q$ which
work equally    well.

   Indeed, $H_q$ can be any $q$ by $q$ matrix of rank $1$
  such that the product of its diagonal entries is  $1/q!$.
  Or, $H_q$ can be the following variant of $E_q$:

  \begin{equation}
  H'_q=D_q\left(\begin{array}{ccccc} 1&-1&0&\cdots&0\\
-1&1&-2&\cdots&0\\
\vdots&\vdots&\vdots&&\vdots\\
(-1)^{q-2}&(-1)^{q-3}&\cdot&\cdots&1-q\\
(-1)^{q-1}&(-1)^{q-2}&\cdot&\cdots&1
\end{array}\right)
\end{equation}

We could change $E_q$ by rescaling all its rows (or all its
columns) by scalars $(r_1,\dots,r_q)$ where $r_1\cdots r_q=1$. But
we do not know of any  significantly different    way to choose
$E_q$.
\end{remark}

\section{ Proof of Proposition  \ref{prop:proj}}
\label{sec:proof_prop_proj}

The aim of this section is to prove Proposition  \ref{prop:proj}
using the results of the last two sections. Putting together
Lemmas \ref{lem:perm_block}  and \ref{lem:HandE} we obtain the two
formulas

\begin{equation}
  \imm_\la\left( \begin{array}{cc}A&0\\ 0&H_q\end{array}
  \right)=\sum_{|\al|=n-q} c^\la_{\al,(q)} \imm_\al(A)
\end{equation}
and
\begin{equation}
  \imm_\la\left( \begin{array}{cc}A&0\\ 0&E_q\end{array}
  \right)=\sum_{|\al|=n-q} c^\la_{\al,(1^q)} \imm_\al(A)
\end{equation}
So proving Proposition  \ref{prop:proj} reduces to proving (I)
$c^\la_{\al,(q)}$ vanishes unless $\al$ is obtained  by removing a
horizontal strip  of size $q$ from $\la$, in which case
$c^\la_{\al,(q)}=1$, and (II) $c^\la_{\al,(1^q)}$ vanishes unless
$\al$ is obtained  by removing a horizontal strip  of size $q$
from $\la$, in which case $c^\la_{\al,(1^q)}=1$.

These statements (I) and (II) are actually familiar facts from the theory
of symmetric group representations and symmetric functions.
The best reference is probably Macdonald's book \cite{ma},
 and so we will explain how to locate these results in his book.

We defined the $c^\la_{\al,\be}$ by the decomposition
\begin{equation}\label{eq:decomp_induced}
\Ind_{\cS_p\times\cS_q}^{\cS_{p+q}}~V_\al\otimes V_\be
=\sum_{|\la|=p+q}c^\la_{\al,\be}V_\la
\end{equation}
But there is a natural vector space isomorphism
$\Psi:\cR\longrightarrow \Lambda$ from the direct sum
$\cR=\oplus_{m\geq 0} \cR^m$ of the character groups of the
symmetric groups $\cS_m$ onto the algebra $\Lambda$ of symmetric
functions in (infinitely many) indeterminates $x_1,x_2,\dots$ with
integral coefficients. This isomorphism $\Psi$ sends the
 character $\chi_\ga$ of $V_\ga$ to the Schur function
$s_\ga=s_\ga(x_1,x_2,\dots)$. In particular
$\Psi(\chi_{(m)})=s_{(m)}$ is the complete homogeneous symmetric
function $h_m$ and $\Psi(\chi_{(1^m)})=s_{(1^m)}$ is the
elementary symmetric function $e_m$. The isomorphism $\Psi$ sends
the character of the induced representation
$\Ind_{\cS_p\times\cS_q}^{\cS_{p+q}}~V_\al\otimes V_\be$ to the
product $s_\al s_\be$ of the corresponding Schur functions
\cite[Chap. I, Sec. 7]{ma}. Thus $\Psi$ transforms
(\ref{eq:decomp_induced}) into the symmetric function formula
\begin{equation}
  s_\al s_\be=\sum_\la c^\la_{\al,\be}s_\la
\end{equation}
 In this context, the $c^\la_{\al,\be}$ are known as
  the Littlewood-Richardson coefficients.

Now computing $c^\la_{\al,(q)}$ and $c^\la_{\al,(1^q)}$ amounts to
 computing the Schur function expansions of $s_\al h_q$ and $s_\al e_q$.
 Macdonald computes these expansions in \cite[Chap. I, Sec. 5,
 (5.16) and (5.17)]{ma}, and he obtains precisely (I) and (II).

\begin{remark}
\label{remark:number} There is a third projection result similar
to \textup{(i)} and \textup{(ii)} in Proposition \ref{prop:proj}.
This result was found by B\"urgisser \cite[Lemma 5.1]{bur1},
\cite[Lemma 7.12, p. 129]{bur_book} and is one of two main tools
he uses to prove $\VNP$-completeness. (A special case was already
used by Hartmann \cite[proof of Lemma 2]{ha}).

The result is that $\Imm_\la$ projects to $\sum_\eta
(-1)^{r(\la,\eta)} \Imm_\eta$ where the sum is over all partitions
$\eta$ obtained by removing from $\la$ a ``skew-hook" $\theta$ of
size $q$ and $r(\la,\eta)$ is one less than the number of rows of
$\theta$. This projection is realized by the formula
\begin{equation}\label{eq:Pq}
\sum_\eta (-1)^{r(\la,\eta)} \imm_\eta(A)=\imm_\la
\left(\begin{array}{cc} A&0\\ 0&P_q\end{array}\right)
\end{equation}
where $P_q$ is the $q$ by $q$ permutation matrix corresponding to
the cycle $(12\cdots q)$.

We note that (\ref{eq:Pq}) can be proven by the same method we
used to prove (\ref{eq:proj_row}) and (\ref{eq:proj_column}).
Indeed, we find that $\imm_\la\left(\begin{array}{cc} A&0\\
0&P_q\end{array}\right)= \sum_{|\al|=n-q} f^\la_\al ~\imm_\al(A)$
where the coefficients $f^\la_\al$ give the Schur function
expansion $s_\al p_q=\sum_{\la} f^\la_\al s_\la$. Here $p_q$ is as
in Remark \ref{rem:Littlewood}. Macdonald computes ${s_\al} p_q$
in \cite[Sec. 3, Example 11]{ma}.
\end{remark}

\section{Proof of Theorem \ref{thm:VNP_compl}}\label{sec:thm1.1}

The aim of this section is to prove Theorem \ref{thm:VNP_compl}.
First Proposition \ref{prop:Im_is_vnp} says that the family
$(\Imlan)$ belongs to $\VNP$. So by Corollary
\ref{cor:VNP_criterion} it is enough to show that if $(\lambn)$
satisfies \textup{(i)} and \textup{(ii)} then the family
$(\Imlan)$ projects to the permanent family $(\Per_n)$. We know by
Corollary \ref{cor:proj} that the polynomial $\Imlan$ projects to
the polynomial $\Per_{k(n)}$  where $k(n)$ is the separation of
$\lambn$. To get the projection result for families, we need to
find a function $t(n)$ such that \textup{(a)} $t(n)$ is
polynomially bounded  and \textup{(b)} the polynomial
$\Imm_{\la^{(t(n))}}$ projects to the polynomial $\Per_n$. Clearly
(b) happens if $k(t(n))\geq n$. But we have in \textup{(ii)} the
growth condition $k(n)=\Omega(n^\delta)$ for some $\delta$, and so
clearly we can find $t(n)$ such that $k(t(n))\geq n$ and
$t(n)=O(n^{1/\delta})$. Thus $t(n)$ satisfies both \textup{(a)}
and \textup{(b)}. This proves that $(\Imlan)$ projects to
$(\Per_n)$. \vskip .16 in

\bibliographystyle{plain}

\end{document}